\documentclass[9pt,%lineno %, onehalfspacing
]{elife}\usepackage[]{graphicx}\usepackage[dvipsnames]{xcolor}
\makeatletter
\def\maxwidth{ %
  \ifdim\Gin@nat@width>\linewidth
    \linewidth
  \else
    \Gin@nat@width
  \fi
}
\makeatother

\definecolor{fgcolor}{rgb}{0.345, 0.345, 0.345}
\newcommand{\hlnum}[1]{\textcolor[rgb]{0.686,0.059,0.569}{#1}}%
\newcommand{\hlcom}[1]{\textcolor[rgb]{0.678,0.584,0.686}{\textit{#1}}}%
\newcommand{\hlopt}[1]{\textcolor[rgb]{0,0,0}{#1}}%
\newcommand{\hlstd}[1]{\textcolor[rgb]{0.345,0.345,0.345}{#1}}%
\newcommand{\hlkwa}[1]{\textcolor[rgb]{0.161,0.373,0.58}{\textbf{#1}}}%
\newcommand{\hlkwb}[1]{\textcolor[rgb]{0.69,0.353,0.396}{#1}}%
\newcommand{\hlkwc}[1]{\textcolor[rgb]{0.333,0.667,0.333}{#1}}%
\newcommand{\hlkwd}[1]{\textcolor[rgb]{0.737,0.353,0.396}{\textbf{#1}}}%

\usepackage{framed}
\makeatletter
\newenvironment{kframe}{%
 \def\at@end@of@kframe{}%
 \ifinner\ifhmode%
  \def\at@end@of@kframe{\end{minipage}}%
  \begin{minipage}{\columnwidth}%
 \fi\fi%
 \def\FrameCommand##1{\hskip\@totalleftmargin \hskip-\fboxsep
 \colorbox{shadecolor}{##1}\hskip-\fboxsep
     \hskip-\linewidth \hskip-\@totalleftmargin \hskip\columnwidth}%
 \MakeFramed {\advance\hsize-\width
   \@totalleftmargin\z@ \linewidth\hsize
   \@setminipage}}%
 {\par\unskip\endMakeFramed%
 \at@end@of@kframe}
\makeatother

\definecolor{shadecolor}{rgb}{.97, .97, .97}
\definecolor{messagecolor}{rgb}{0, 0, 0}
\definecolor{warningcolor}{rgb}{1, 0, 1}
\definecolor{errorcolor}{rgb}{1, 0, 0}
\newenvironment{knitrout}{}{} % an empty environment to be redefined in TeX

\usepackage{alltt}
\usepackage[T1]{fontenc}
\usepackage[utf8]{inputenc}
\usepackage[english]{babel}
\usepackage[dvipsnames]{xcolor}
\usepackage{tikz} % to draw schematics
\usepackage{doi}
\usetikzlibrary{decorations.pathreplacing,calligraphy} % for tikz curly braces
\usepackage{todonotes}
\usepackage{boxedminipage}
\usepackage{nameref}
\usepackage{caption}
\definecolor{col1}{HTML}{140e09}
\definecolor{col2}{HTML}{4daf4a}

\title{Replication of ``null results'' -- Absence of evidence or evidence of
  absence?}

\author[1*\authfn{1}]{Samuel Pawel}
\author[1\authfn{1}]{Rachel Heyard}
\author[1]{Charlotte Micheloud}
\author[1]{Leonhard Held}
\affil[1]{Epidemiology, Biostatistics and Prevention Institute, Center for Reproducible Science, University of Zurich, Switzerland}

\corr{samuel.pawel@uzh.ch}{SP}

\contrib[\authfn{1}]{Contributed equally}

\DeclareMathOperator{\Nor}{N} % Normal
\newcommand{\given}{\,\vert\,} % Given
\DeclareMathOperator{\BF}{BF} % Bayes factor

\IfFileExists{upquote.sty}{\usepackage{upquote}}{}
\begin{document}
\maketitle

\begin{abstract}
  In several large-scale replication projects, statistically non-significant
  results in both the original and the replication study have been interpreted
  as a ``replication success''. Here we discuss the logical problems with this
  approach: Non-significance in both studies does not ensure that the studies
  provide evidence for the absence of an effect and ``replication success'' can
  virtually always be achieved if the sample sizes are small enough. In
  addition, the relevant error rates are not controlled. We show how methods,
  such as equivalence testing and Bayes factors, can be used to adequately
  quantify the evidence for the absence of an effect and how they can be applied
  in the replication setting. Using data from the Reproducibility Project:
  Cancer Biology, the Experimental Philosophy Replicability Project, and the
  Reproducibility Project: Psychology we illustrate that many original and
  replication studies with ``null results'' are in fact inconclusive. We
  conclude that it is important to also replicate studies with statistically
  non-significant results, but that they should be designed, analyzed, and
  interpreted appropriately.
\end{abstract}

% \emph{Keywords}: Bayesian hypothesis testing, equivalence testing,
% meta-research, null hypothesis, replication success

\section{Introduction}

\textit{Absence of evidence is not evidence of absence} -- the title of the 1995
paper by Douglas Altman and Martin Bland has since become a mantra in the
statistical and medical literature \citep{Altman1995}. Yet, the misconception
that a statistically non-significant result indicates evidence for the absence
of an effect is unfortunately still widespread \citep{Greenland2011, Makin2019}.
Such a ``null result'' -- typically characterized by a \textit{p}-value
$p > 0.05$ for the null hypothesis of an absent effect -- may also occur if an
effect is actually present. For example, if the sample size of a study is chosen
to detect an assumed effect with a power of 80\%, null results will incorrectly
occur 20\% of the time when the assumed effect is actually present. If the power
of the study is lower, null results will occur more often. In general, the lower
the power of a study, the greater the ambiguity of a null result. To put a null
result in context, it is therefore critical to know whether the study was
adequately powered and under what assumed effect the power was calculated
\citep{Hoenig2001, Greenland2012}. However, if the goal of a study is to
explicitly quantify the evidence for the absence of an effect, more appropriate
methods designed for this task, such as equivalence testing
\citep{Wellek2010,Lakens2017, Senn2021} or Bayes factors \citep{Kass1995,
  Goodman1999,Goodman2005,Dienes2014,Keysers2020}, should be used from the
outset.

The interpretation of null results becomes even more complicated in the setting
of replication studies. In a replication study, researchers attempt to repeat an
original study as closely as possible in order to assess whether consistent
results can be obtained with new data \citep{NSF2019}. In the last decade,
various large-scale replication projects have been conducted in diverse fields,
from the biomedical to the social sciences \citep[among
others]{Prinz2011,Begley2012,Klein2014,Opensc2015,Camerer2016,Camerer2018,Klein2018,Cova2018,Errington2021}.
The majority of these projects reported alarmingly low replicability rates
across a broad spectrum of criteria for quantifying replicability. While most of
these projects restricted their focus on original studies with statistically
significant results (``positive results''), the \emph{Reproducibility Project:
  Cancer Biology} \citep[RPCB,][]{Errington2021}, the \emph{Experimental
  Philosophy Replicability Project} \citep[EPRP,][]{Cova2018}, and the
\emph{Reproducibility Project: Psychology} \citep[RPP,][]{Opensc2015} also
attempted to replicate some original studies with null results -- either
non-significant or interpreted as showing no evidence for a meaningful effect by
the original authors.

Although the EPRP and RPP interpreted non-significant results in both original
and replication study as a ``replication success'' for some individual
replications (see, for example, the replication of \citet[replication report:
\url{https://osf.io/wcm7n}]{McCann2005} or the replication of \citet[replication
report: \url{https://osf.io/9xt25}]{Ranganath2008}),
% and \url{https://osf.io/fkcn5})
they excluded the original null results in the calculation of an overall
replicability rate based on significance. In contrast, the RPCB explicitly
defined null results in both the original and the replication study as a
criterion for ``replication success''. According to this ``non-significance''
criterion, 11/15 = 73\% replications of original null
effects were successful. Four additional criteria were used to provide a more
nuanced assessment of replication success for original null results: (i) whether
the original effect estimate was included in the 95\% confidence interval of the
replication effect estimate (success rate 11/15 = 73\%), (ii) whether the replication effect estimate was included in the 95\%
confidence interval of the original effect estimate (success rate 12/15 =
80\%), (iii) whether the replication effect estimate
was included in the 95\% prediction interval based on the original effect
estimate (success rate 12/15 = 80\%), (iv) and whether
the \textit{p}-value obtained from combining the original and replication effect
estimate with a meta-analysis was non-significant (success rate 10/15 =
67\%). Criteria (i) to (iii) are useful for assessing
compatibility in effect estimates between the original and the replication
study. Their suitability has been extensively discussed in the literature. The
prediction interval criterion (iii) or equivalent criteria (e.g., the $Q$-test)
are usually recommended because they account for the uncertainty from both
studies and have adequate error rates when the true effect sizes are the same
\citep{Patil2016, Mathur2020, Schauer2021}.

While the effect estimate criteria (i) to (iii) can be applied regardless of
whether or not the original study was non-significant, the ``meta-analytic
non-significance'' criterion (iv) and the aforementioned non-significance
criterion refer specifically to original null results. We believe that there are
several logical problems with both, and that it is important to highlight and
address them, especially since the non-significance criterion has already been
used in three replication projects without much scrutiny. It is crucial to note
that it is not our intention to diminish the enormously important contributions
of the RPCB, the EPRP, and the RPP, but rather to build on their work and
provide recommendations for ongoing and future replication projects
\citep[e.g.,][]{Amaral2019, Murphy2022}.

The logical problems with the non-significance criterion are as follows: First,
if the original study had low statistical power, a non-significant result is
highly inconclusive and does not provide evidence for the absence of an effect.
It is then unclear what exactly the goal of the replication should be -- to
replicate the inconclusiveness of the original result? On the other hand, if the
original study was adequately powered, a non-significant result may indeed
provide some evidence for the absence of an effect when analyzed with
appropriate methods, so that the goal of the replication is clearer. However,
the criterion by itself does not distinguish between these two cases. Second,
with this criterion researchers can virtually always achieve replication success
by conducting a replication study with a very small sample size, such that the
\textit{p}-value is non-significant and the result is inconclusive. This is
because the null hypothesis under which the \textit{p}-value is computed is
misaligned with the goal of inference, which is to quantify the evidence for the
absence of an effect. Third, the criterion does not control the error of falsely
claiming the absence of an effect at a predetermined rate. This is in contrast
to the standard criterion for replication success, which requires significance
from both studies \citep[also known as the two-trials rule, see Section 12.2.8
in][]{Senn2021}, and ensures that the error of falsely claiming the presence of
an effect is controlled at a rate equal to the squared significance level (for
example, 5\% $\times$ 5\% = 0.25\% for a 5\% significance level). The
non-significance criterion may be intended to complement the two-trials rule for
null results. However, it fails to do so in this respect, which may be required
by regulators and funders. These logical problems are equally applicable to the
meta-analytic non-significance criterion.

In the following, we present two principled approaches for analyzing replication
studies of null results -- frequentist equivalence testing and Bayesian
hypothesis testing -- that can address the limitations of the non-significance
criterion. We use the null results replicated in the RPCB, RPP, and EPRP to
illustrate the problems of the non-significance criterion and how they can be
addressed. We conclude the paper with practical recommendations for analyzing
replication studies of original null results, including simple R code for
applying the proposed methods.

\section{Null results from the Reproducibility Project: Cancer Biology}
\label{sec:rpcb}

\begin{figure}[!htb]
\begin{knitrout}
\definecolor{shadecolor}{rgb}{0.969, 0.969, 0.969}\color{fgcolor}
\includegraphics[width=\maxwidth]{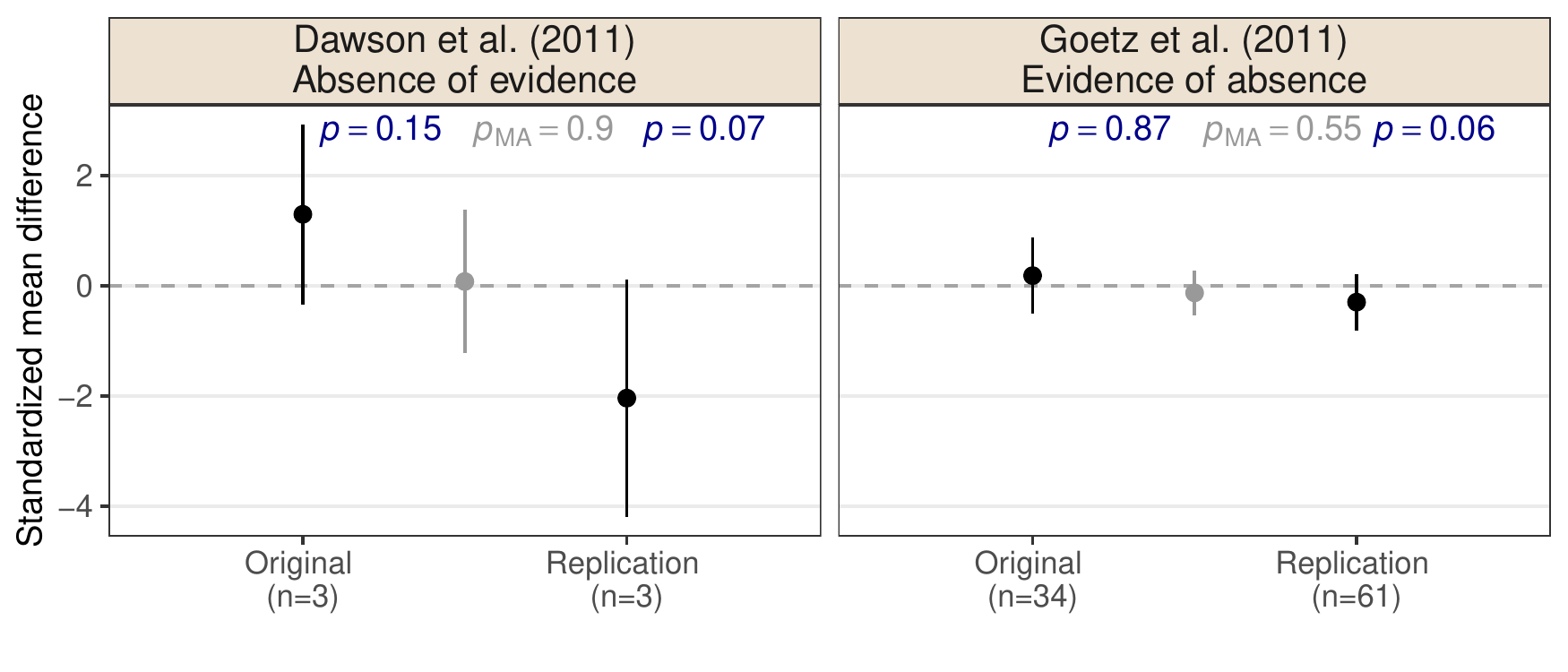} 
\end{knitrout}
\caption{\label{fig:2examples} Two examples of original and replication study
  pairs which meet the non-significance replication success criterion from the
  Reproducibility Project: Cancer Biology \citep{Errington2021}. Shown are
  standardized mean difference effect estimates with 95\% confidence intervals, sample sizes \textit{n}, and two-sided
  \textit{p}-values \textit{p} for the null hypothesis that the effect is
  absent. Effect estimate, 95\% confidence interval, and \textit{p}-value from a
  fixed-effect meta-analysis \textit{p}\textsubscript{MA} of original and
  replication study are shown in gray.}
\end{figure}

Figure~\ref{fig:2examples} shows effect estimates on standardized mean
difference (SMD) scale with 95\% confidence
intervals from two RPCB study pairs. In both study pairs, the original and
replication studies are ``null results'' and therefore meet the non-significance
criterion for replication success (the two-sided \textit{p}-values are greater
than 0.05 in both the original and the replication study). The same is true when
applying the meta-analytic non-significance criterion (the two-sided
\textit{p}-values of the meta-analyses \textit{p}\textsubscript{MA} are greater
than 0.05). However, intuition would suggest that the conclusions in the two
pairs are very different.

The original study from \citet{Dawson2011} and its replication both show large
effect estimates in magnitude, but due to the very small sample sizes, the
uncertainty of these estimates is large, too. With such low sample sizes, the
results seem inconclusive. In contrast, the effect estimates from
\citet{Goetz2011} and its replication are much smaller in magnitude and their
uncertainty is also smaller because the studies used larger sample sizes.
Intuitively, the results seem to provide more evidence for a zero (or negligibly
small) effect. While these two examples show the qualitative difference between
absence of evidence and evidence of absence, we will now discuss how the two can
be quantitatively distinguished.

\section{Methods for assessing replicability of null results}
\label{sec:methods}
There are both frequentist and Bayesian methods that can be used for assessing
evidence for the absence of an effect. \citet{Anderson2016} provide an excellent
summary in the context of replication studies in psychology. We now briefly
discuss two possible approaches -- frequentist equivalence testing and Bayesian
hypothesis testing -- and their application to the RPCB, EPRP, and RPP data.

\subsection{Frequentist equivalence testing}
\begin{figure}[!htb]
  \begin{center}
    \begin{tikzpicture}[ultra thick]
      \draw[stealth-stealth] (0,0) -- (6,0);
      \node[text width=4.5cm, align=center] at (3,-1) {Effect size};
      \draw (2,0.2) -- (2,-0.2) node[below]{$-\Delta$};
      \draw (3,0.2) -- (3,-0.2) node[below]{$0$};
      \draw (4,0.2) -- (4,-0.2) node[below]{$+\Delta$};

      \node[text width=5cm, align=left] at (0,1) {\textbf{Equivalence}};
      \draw [draw={col1},decorate,decoration={brace,amplitude=5pt}]
      (2.05,0.75) -- (3.95,0.75) node[midway,yshift=1.5em]{\textcolor{col1}{$H_1$}};
      \draw [draw={col2},decorate,decoration={brace,amplitude=5pt,aspect=0.6}]
      (0,0.75) -- (1.95,0.75) node[pos=0.6,yshift=1.5em]{\textcolor{col2}{$H_0$}};
      \draw [draw={col2},decorate,decoration={brace,amplitude=5pt,aspect=0.4}]
      (4.05,0.75) -- (6,0.75) node[pos=0.4,yshift=1.5em]{\textcolor{col2}{$H_0$}};

      \node[text width=5cm, align=left] at (0,2.15) {\textbf{Superiority}\\(two-sided)};
      \draw [decorate,decoration={brace,amplitude=5pt}]
      (3,2) -- (3,2) node[midway,yshift=1.5em]{\textcolor{col2}{$H_0$}};
      \draw[col2] (3,1.95) -- (3,2.2);
      \draw [draw={col1},decorate,decoration={brace,amplitude=5pt,aspect=0.6}]
      (0,2) -- (2.95,2) node[pos=0.6,yshift=1.5em]{\textcolor{col1}{$H_1$}};
      \draw [draw={col1},decorate,decoration={brace,amplitude=5pt,aspect=0.4}]
      (3.05,2) -- (6,2) node[pos=0.4,yshift=1.5em]{\textcolor{col1}{$H_1$}};

      \node[text width=5cm, align=left] at (0,3.45) {\textbf{Superiority}\\(one-sided)};
      \draw [draw={col1},decorate,decoration={brace,amplitude=5pt,aspect=0.4}]
      (3.05,3.25) -- (6,3.25) node[pos=0.4,yshift=1.5em]{\textcolor{col1}{$H_1$}};
      \draw [draw={col2},decorate,decoration={brace,amplitude=5pt,aspect=0.6}]
      (0,3.25) -- (3,3.25) node[pos=0.6,yshift=1.5em]{\textcolor{col2}{$H_0$}};

      \draw [dashed] (2,0) -- (2,0.75);
      \draw [dashed] (4,0) -- (4,0.75);
      \draw [dashed] (3,0) -- (3,0.75);
      \draw [dashed] (3,1.5) -- (3,1.9);
      \draw [dashed] (3,2.8) -- (3,3.2);
    \end{tikzpicture}
  \end{center}
  \caption{Null hypothesis ($H_0$) and alternative hypothesis ($H_1$) for
    superiority and equivalence tests (with equivalence margin $\Delta > 0$).}
  \label{fig:hypotheses}
\end{figure}
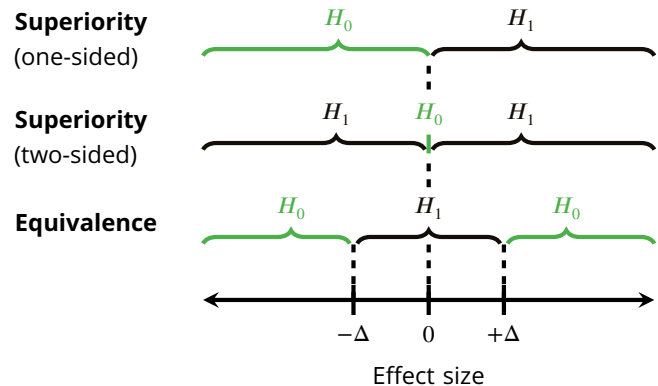

Equivalence testing was developed in the context of clinical trials to assess
whether a new treatment -- typically cheaper or with fewer side effects than the
established treatment -- is practically equivalent to the established treatment
\citep{Wellek2010,Lakens2017}. The method can also be used to assess whether an
effect is practically equivalent to an absent effect, usually zero. Using
equivalence testing as a way to put non-significant results into context has
been suggested by several authors \citep{Hauck1986, Campbell2018}. The main
challenge is to specify the margin $\Delta > 0$ that defines an equivalence
range $[-\Delta, +\Delta]$ in which an effect is considered as absent for
practical purposes. The goal is then to reject the null hypothesis that the true
effect is outside the equivalence range. This is in contrast to the usual null
hypotheses of superiority tests which state that the effect is zero or smaller
than zero, see Figure~\ref{fig:hypotheses} for an illustration.

To ensure that the null hypothesis is falsely rejected at most
$\alpha \times 100\%$ of the time, the standard approach is to declare
equivalence if the $(1-2\alpha)\times 100\%$ confidence interval for the effect
is contained within the equivalence range, for example, a 90\% confidence
interval for $\alpha = 0.05$ \citep{Westlake1972}. This procedure is equivalent
to declaring equivalence when two one-sided tests (TOST) for the null hypotheses
of the effect being greater/smaller than $+\Delta$ and $-\Delta$, are both
significant at level $\alpha$ \citep{Schuirmann1987}. A quantitative measure of
evidence for the absence of an effect is then given by the maximum of the two
one-sided \textit{p}-values -- the TOST \textit{p}-value \citep[section
4.4]{Greenland2023}. In case a dichotomous replication success criterion for
null results is desired, it is natural to require that both the original and the
replication TOST \textit{p}-values are smaller than some level $\alpha$
(conventionally $\alpha = 0.05$). Equivalently, the criterion would require the
$(1-2\alpha)\times 100\%$ confidence intervals of the original and the
replication to be included in the equivalence region. In contrast to the
non-significance criterion, this criterion controls the error of falsely
claiming replication success at level $\alpha^{2}$ when there is a true effect
outside the equivalence margin, thus complementing the usual two-trials rule in
drug regulation \citep[Section 12.2.8]{Senn2021}.

\begin{figure}
  \begin{fullwidth}
\begin{knitrout}
\definecolor{shadecolor}{rgb}{0.969, 0.969, 0.969}\color{fgcolor}
\includegraphics[width=\maxwidth]{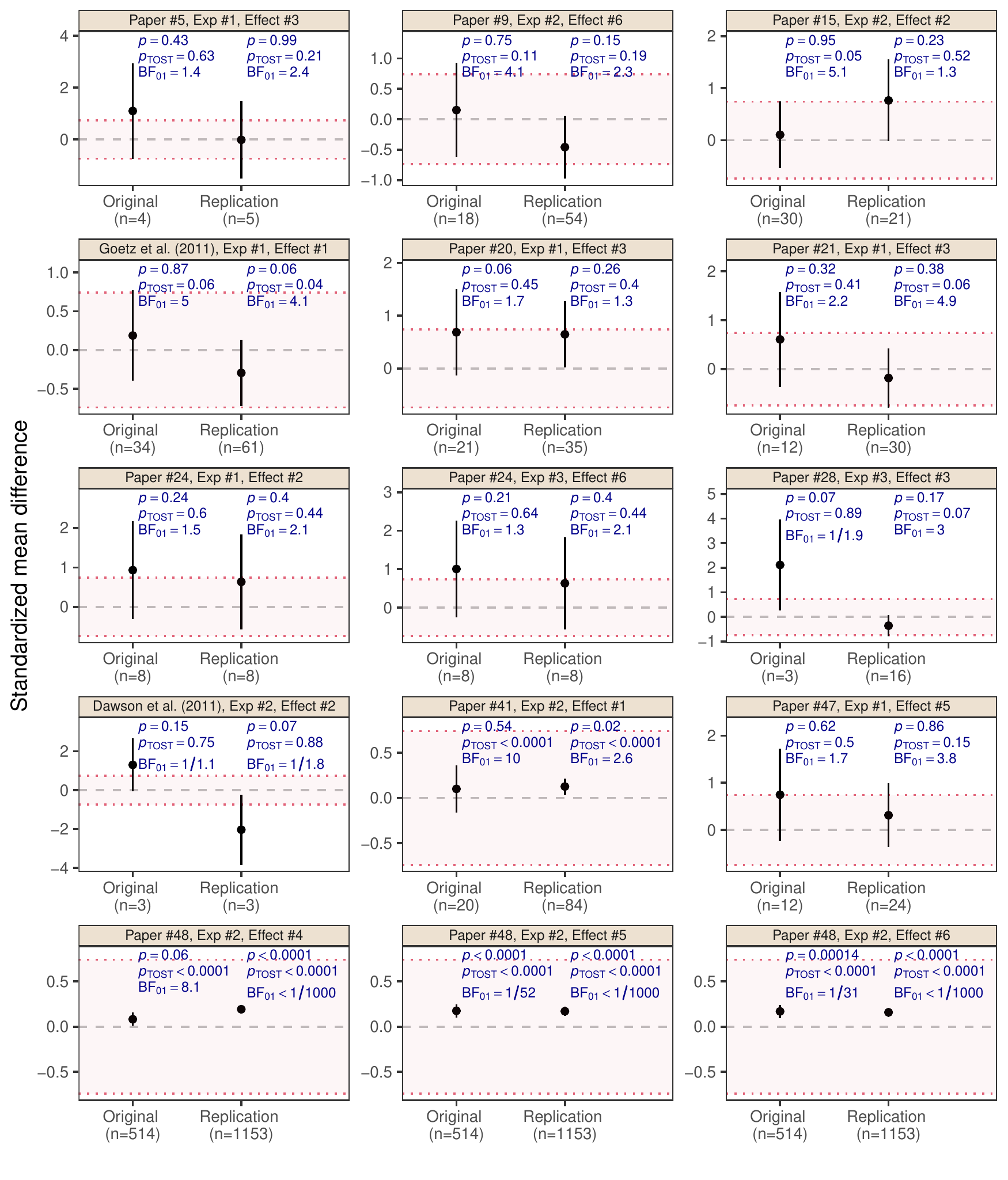} 
\end{knitrout}
\caption{Effect estimates on standardized mean difference (SMD) scale with
  90\% confidence interval for the 15 ``null
  results'' and their replication studies from the Reproducibility Project:
  Cancer Biology \citep{Errington2021}. The title above each plot indicates the
  original paper, experiment and effect numbers. Two original effect estimates
  from original paper 48 were statistically significant at $p < 0.05$, but were
  interpreted as null results by the original authors and therefore treated as
  null results by the RPCB. The two examples from Figure~\ref{fig:2examples} are
  indicated in the plot titles. The dashed gray line represents the value of no
  effect ($\text{SMD} = 0$), while the dotted red lines represent the
  equivalence range with a margin of $\Delta = 0.74$, classified as
  ``liberal'' by \citet[Table 1.1]{Wellek2010}. The \textit{p}-value
  $p_{\text{TOST}}$ is the maximum of the two one-sided \textit{p}-values for
  the null hypotheses of the effect being greater/less than $+\Delta$ and
  $-\Delta$, respectively. The Bayes factor $\BF_{01}$ quantifies the evidence
  for the null hypothesis $H_{0} \colon \text{SMD} = 0$ against the alternative
  $H_{1} \colon \text{SMD} \neq 0$ with normal unit-information prior assigned
  to the SMD under $H_{1}$.}
\label{fig:nullfindings}
\end{fullwidth}
\end{figure}

Returning to the RPCB data, Figure~\ref{fig:nullfindings} shows the standardized
mean difference effect estimates with 90\%
confidence intervals for all 15 effects which were treated as null results by
the RPCB.\footnote{There are four original studies with null effects for which
  two or three ``internal'' replication studies were conducted, leading in total
  to 20 replications of null effects. As done in the RPCB main analysis
  \citep{Errington2021}, we aggregated their SMD estimates into a single SMD
  estimate with fixed-effect meta-analysis and recomputed the replication
  \textit{p}-value based on a normal approximation. For the original studies and
  the single replication studies we report the SMD estimates and
  \textit{p}-values as provided by the RPCB.} Most of them showed
non-significant \textit{p}-values ($p > 0.05$) in the original study. It is
noteworthy, however, that two effects from the second experiment of the original
paper 48 were regarded as null results despite their statistical significance.
According to the non-significance criterion (requiring $p > 0.05$ in original
and replication study), there are 11 ``successes'' out of
total 15 null effects, as reported in Table 1
from~\citet{Errington2021}.

We will now apply equivalence testing to the RPCB data. The dotted red lines in
Figure~\ref{fig:nullfindings} represent an equivalence range for the margin
$\Delta = 0.74$, which \citet[Table  1.1]{Wellek2010} classifies as
``liberal''. However, even with this generous margin, only
4 of the 15 study pairs are able to
establish replication success at the 5\% level, in the sense that both the
original and the replication 90\% confidence interval fall within the
equivalence range (or, equivalently, that their TOST \textit{p}-values are
smaller than 0.05). For the remaining 11
studies, the situation remains inconclusive and there is no evidence for the
absence or the presence of the effect. For instance, the previously discussed
example from \citet{Goetz2011} marginally fails the criterion
($p_{\text{TOST}} = 0.06$ in the original study and
$p_{\text{TOST}} = 0.04$ in the replication), while the
example from \citet{Dawson2011} is a clearer failure
($p_{\text{TOST}} = 0.75$ in the original study and
$p_{\text{TOST}} = 0.88$ in the replication) as both
effect estimates even lie outside the equivalence margin.

The post-hoc specification of equivalence margins is controversial. Ideally, the
margin should be specified on a case-by-case basis in a pre-registered protocol
before the studies are conducted by researchers familiar with the subject
matter. In the social and medical sciences, the conventions of \citet{Cohen1992}
are typically used to classify SMD effect sizes ($\text{SMD} = 0.2$ small,
$\text{SMD} = 0.5$ medium, $\text{SMD} = 0.8$ large). While effect sizes are
typically larger in preclinical research, it seems unrealistic to specify
margins larger than $1$ on SMD scale to represent effect sizes that are absent
for practical purposes. It could also be argued that the chosen margin
$\Delta = 0.74$ is too lax compared to margins commonly used in
clinical research \citep{Lange2005}. We therefore report a sensitivity analysis
regarding the choice of the margin in Figure~\ref{fig:sensitivity} in
Appendix~A. This analysis shows that for realistic margins between $0$ and $1$,
the proportion of replication successes remains below 50\% for the conventional
$\alpha = 0.05$ level. To achieve a success rate of 11/15 =
73\%, as was achieved with the non-significance
criterion from the RPCB, unrealistic margins of $\Delta > 2$ are required.

Appendix~B shows similar equivalence test analyses for the four study pairs with
original null results from the RPP and EPRP. Three study pair results turn out
to be inconclusive due to the large uncertainty around their effect estimates.

\subsection{Bayesian hypothesis testing}
\begin{sloppypar}
The distinction between absence of evidence and evidence of absence is naturally
built into the Bayesian approach to hypothesis testing. A central measure of
evidence is the Bayes factor \citep{Kass1995,
  Goodman1999,Dienes2014,Keysers2020}, which is the updating factor of the prior
odds to the posterior odds of the null hypothesis $H_{0}$ versus the alternative
hypothesis~$H_{1}$
\begin{align*}
\underbrace{\frac{\Pr(H_{0}~\mathrm{given}~\mathrm{data})}{\Pr(H_{1}~\mathrm{given}~
  \mathrm{data})}}_{\mathrm{Posterior~odds}}
  =  \underbrace{\frac{\Pr(H_{0})}{\Pr(H_{1})}}_{\mathrm{Prior~odds}}
  \times \underbrace{\frac{\Pr(\mathrm{data}~\mathrm{given}~ H_{0})}{\Pr(\mathrm{data}
  ~\mathrm{given}~H_{1})}}_{\mathrm{Bayes~factor}~\BF_{01}}.
\end{align*}
The Bayes factor $\BF_{01}$ quantifies how much the observed data have increased
or decreased the probability $\Pr(H_{0})$ of the null hypothesis relative to the
probability $\Pr(H_{1})$ of the alternative. As such, Bayes factor are direct
measures of evidence for the null hypothesis, in contrast to \textit{p}-values,
which are only indirect measures of evidence as they are computed under the
assumption that the null hypothesis is true \citep{Held2018}. If the null
hypothesis states the absence of an effect, a Bayes factor greater than one
(\mbox{$\BF_{01} > 1$}) indicates evidence for the absence of the effect and a
Bayes factor smaller than one indicates evidence for the presence of the effect
(\mbox{$\BF_{01} < 1$}), whereas a Bayes factor not much different from one
indicates absence of evidence for either hypothesis
(\mbox{$\BF_{01} \approx 1$}). Bayes factors are quantitative summaries of the
evidence provided by the data in favor of the null hypothesis as opposed to the
alternative hypothesis. If a dichotomous criterion for successful replication of
a null result is desired, it seems natural to require a Bayes factor larger than
some level $\gamma > 1$ from both studies, for example, $\gamma = 3$ or
$\gamma = 10$ which are conventional levels for ``substantial'' and ``strong''
evidence, respectively \citep{Jeffreys1961}. In contrast to the non-significance
criterion, this criterion provides a genuine measure of evidence that can
distinguish absence of evidence from evidence of absence.
\end{sloppypar}

% When the observed data are dichotomized into positive (\mbox{$p < 0.05$}) or
% null results (\mbox{$p > 0.05$}), the Bayes factor based on a null result is the
% probability of observing \mbox{$p > 0.05$} when the effect is indeed absent
% (which is $95\%$) divided by the probability of observing $p > 0.05$ when the
% effect is indeed present (which is one minus the power of the study). For
% example, if the power is $90\%$, we have
% \mbox{$\BF_{01} = 95\%/10\% = round(0.95/0.1, 2)$} indicating almost ten
% times more evidence for the absence of the effect than for its presence. On the
% other hand, if the power is only $50\%$, we have
% \mbox{$\BF_{01} = 95\%/50\% = round(0.95/0.5,2)$} indicating only
% slightly more evidence for the absence of the effect. This example also
% highlights
The main challenge with Bayes factors is the specification of the effect under
the alternative hypothesis $H_{1}$. The assumed effect under $H_{1}$ is directly
related to the Bayes factor,
% is directly related to the power of the study,
and researchers who assume different effects will end up with different Bayes
factors. Instead of specifying a single effect, one therefore typically
specifies a ``prior distribution'' of plausible effects. Importantly, the prior
distribution, like the equivalence margin, should be determined by researchers
with subject knowledge and before the data are collected.

% In practice, the observed data should not be dichotomized into positive or null
% results, as this leads to a loss of information. Therefore,
To compute the Bayes factors for the RPCB null results, we used the observed
effect estimates as the data and assumed a normal sampling distribution for them
\citep{Dienes2014}, as typically done in a meta-analysis. The Bayes factors
$\BF_{01}$ shown in Figure~\ref{fig:nullfindings} then quantify the evidence for
the null hypothesis of no effect % ($H_{0} \colon \text{SMD} = 0$)
against the
alternative hypothesis that there is an effect
% ($H_{1} \colon \text{SMD} \neq 0$)
using a normal ``unit-information'' prior distribution \citep{Kass1995b} for the
effect size under the alternative $H_{1}$, see Appendix~C for further details on
the calculation of these Bayes factors. We see that in most cases there is no
substantial evidence for either the absence or the presence of an effect, as
with the equivalence tests. For instance, with a lenient Bayes factor threshold
of $3$, only 1 of the 15 replications are
successful, in the sense of having $\BF_{01} > 3$ in both the original and the
replication study. The Bayes factors for the two previously discussed examples
are consistent with our intuitions -- in the \citet{Goetz2011} example there is
indeed substantial evidence for the absence of an effect
($\BF_{01} = 5$ in the original study and
$\BF_{01} = 4.1$ in the replication), while in the
\citet{Dawson2011} example there is even anecdotal evidence for the
\emph{presence} of an effect, though the Bayes factors are very close to one due
to the small sample sizes ($\BF_{01} = 1/1.1$ in the
original study and $\BF_{01} = 1/1.8$ in the replication).

As with the equivalence margin, the choice of the prior distribution for the SMD
under the alternative $H_{1}$ is debatable. The normal unit-information prior
seems to be a reasonable default choice, as it implies that small to large
effects are plausible under the alternative, but other normal priors with
smaller/larger standard deviations could have been considered to make the test
more sensitive to smaller/larger true effect sizes. The sensitivity analysis in
Appendix~A therefore also includes an analysis on the effect of varying prior
standard deviations and the Bayes factor thresholds. However, again, to achieve
replication success for a larger proportion of replications than the observed
1/15 = 7\%,
unreasonably large prior standard deviations have to be specified.

Of note, among the 15 RPCB null results, there are three interesting
cases (the three effects from original paper 48 by \citealp{Lin2012} and its
replication by \citealp{Lewis2018}) where the Bayes factor is qualitatively
different from the equivalence test, revealing a fundamental difference between
the two approaches. The Bayes factor is concerned with testing whether the
effect is \emph{exactly zero}, whereas the equivalence test is concerned with
whether the effect is within an \emph{interval around zero}. Due to the very
large sample size in the original study (\textit{n} = 514) and
the replication (\textit{n} = 1'153),
the data are incompatible with an exactly zero effect, but compatible with
effects within the equivalence range. Apart from this example, however, both
approaches lead to the same qualitative conclusion -- most RPCB null results are
highly ambiguous.

Appendix~B also shows Bayes factor analyses for the four study pairs with
original null results from the RPP and EPRP. In contrast to the RPCB results,
most Bayes factors indicate non-anecdotal evidence for a null effect in cases
where the non-signiﬁcance criterion was met, possibly because of the larger
sample sizes and smaller effects in these ﬁelds.

\begin{table}[!htb]
  \centering \small
  \caption*{Box 1: Recommendations for the analysis of replication studies of
    original null results. Calculations are based on effect estimates
    $\hat{\theta}_{i}$ with standard errors $\sigma_{i}$ from an original study
    ($i = o$) and its replication ($i = r$). Both effect estimates are assumed
    to be normally distributed around the true effect size $\theta$ with known
    variance $\sigma^{2}_{i}$. The effect size $\theta_{0}$ represents the value
    of no effect, typically $\theta_{0} = 0$.}
  \label{box:recommendations}
  \begin{boxedminipage}[c]{\linewidth}
    \small
\textbf{Equivalence test}
      \begin{enumerate}
        \item Specify a margin $\Delta > 0$ that defines an equivalence range
              $[\theta_{0} - \Delta, \theta_{0} + \Delta]$ in which effects are
              considered absent for practical purposes.
        \item Compute the TOST \textit{p}-values for original ($i = o$) and
              replication ($i = r$) data
              $$p_{\text{TOST},i}
              = \max\left\{\Phi\left(\frac{\hat{\theta}_{i} - \theta_{0} - \Delta}{\sigma_{i}}\right),
              1 - \Phi\left(\frac{\hat{\theta}_{i} - \theta_{0} + \Delta}{\sigma_{i}}\right)\right\},$$
              with $\Phi(\cdot)$ the cumulative distribution function of the
              standard normal distribution.
\begin{minipage}[c]{0.95\linewidth}
\begin{knitrout}\small
\definecolor{shadecolor}{rgb}{0.969, 0.969, 0.969}\color{fgcolor}\begin{kframe}
\begin{alltt}
\hlcom{## R function to compute TOST p-value based on effect estimate, standard error,}
\hlcom{## null value (default is 0), and equivalence margin specified in step 1.}
\hlstd{pTOST} \hlkwb{<-} \hlkwa{function}\hlstd{(}\hlkwc{estimate}\hlstd{,} \hlkwc{se}\hlstd{,} \hlkwc{null} \hlstd{=} \hlnum{0}\hlstd{,} \hlkwc{margin}\hlstd{) \{}
    \hlstd{p1} \hlkwb{<-} \hlkwd{pnorm}\hlstd{(}\hlkwc{q} \hlstd{= (estimate} \hlopt{-} \hlstd{null} \hlopt{-} \hlstd{margin)} \hlopt{/} \hlstd{se)}
    \hlstd{p2} \hlkwb{<-} \hlnum{1} \hlopt{-} \hlkwd{pnorm}\hlstd{(}\hlkwc{q} \hlstd{= (estimate} \hlopt{-} \hlstd{null} \hlopt{+} \hlstd{margin)} \hlopt{/} \hlstd{se)}
    \hlstd{p} \hlkwb{<-} \hlkwd{pmax}\hlstd{(p1, p2)}
    \hlkwd{return}\hlstd{(p)}
\hlstd{\}}
\end{alltt}
\end{kframe}
\end{knitrout}
\end{minipage}
        \item Declare replication success at level $\alpha$ if
              $p_{\text{TOST},o} \leq \alpha$ and $p_{\text{TOST},r} \leq \alpha$,
              conventionally $\alpha = 0.05$.
        \item Perform a sensitivity analysis with respect to the margin
              $\Delta$. For example, visualize the TOST \textit{p}-values for
              different margins to assess the robustness of the conclusions. \\
      \end{enumerate}

      \textbf{Bayes factor}
      \begin{enumerate}
        \item Specify a prior distribution for the effect size $\theta$ that
              represents plausible values under the alternative hypothesis that
              there is an effect ($H_{1}\colon \theta \neq \theta_{0})$. For
              example, specify the mean $m$ and standard deviation $s$ of a normal
              distribution $\theta \given H_{1} \sim \Nor(m, s^{2})$.
        \item Compute the Bayes factors contrasting
              $H_{0} \colon \theta = \theta_{0}$ to
              $H_{1} \colon \theta \neq \theta_{0}$ for original ($i = o$) and
              replication ($i = r$) data. Assuming a normal prior distribution,
              % $\theta \given H_{1} \sim \Nor(m ,v)$,
              the Bayes factor is
              $$\BF_{01,i}
              = \sqrt{1 + \frac{s^{2}}{\sigma^{2}_{i}}} \, \exp\left[-\frac{1}{2} \left\{\frac{(\hat{\theta}_{i} -
              \theta_{0})^{2}}{\sigma^{2}_{i}} - \frac{(\hat{\theta}_{i} - m)^{2}}{\sigma^{2}_{i} + s^2}
              \right\}\right].$$
\begin{minipage}[c]{0.95\linewidth}
\begin{knitrout}\small
\definecolor{shadecolor}{rgb}{0.969, 0.969, 0.969}\color{fgcolor}\begin{kframe}
\begin{alltt}
\hlcom{## R function to compute Bayes factor based on effect estimate, standard error,}
\hlcom{## null value (default is 0), prior mean (default is null value), and prior}
\hlcom{## standard deviation specified in step 1.}
\hlstd{BF01} \hlkwb{<-} \hlkwa{function}\hlstd{(}\hlkwc{estimate}\hlstd{,} \hlkwc{se}\hlstd{,} \hlkwc{null} \hlstd{=} \hlnum{0}\hlstd{,} \hlkwc{priormean} \hlstd{= null,} \hlkwc{priorsd}\hlstd{) \{}
    \hlstd{bf} \hlkwb{<-} \hlkwd{sqrt}\hlstd{(}\hlnum{1} \hlopt{+} \hlstd{priorsd}\hlopt{^}\hlnum{2}\hlopt{/}\hlstd{se}\hlopt{^}\hlnum{2}\hlstd{)} \hlopt{*} \hlkwd{exp}\hlstd{(}\hlopt{-}\hlnum{0.5} \hlopt{*} \hlstd{((estimate} \hlopt{-} \hlstd{null)}\hlopt{^}\hlnum{2} \hlopt{/} \hlstd{se}\hlopt{^}\hlnum{2} \hlopt{-}
            \hlstd{(estimate} \hlopt{-} \hlstd{priormean)}\hlopt{^}\hlnum{2} \hlopt{/} \hlstd{(se}\hlopt{^}\hlnum{2} \hlopt{+} \hlstd{priorsd}\hlopt{^}\hlnum{2}\hlstd{)))}
    \hlkwd{return}\hlstd{(bf)}
\hlstd{\}}
\end{alltt}
\end{kframe}
\end{knitrout}
\end{minipage}
        \item Declare replication success at level $\gamma > 1$ if
              $\BF_{01,o} \geq \gamma$ and $\BF_{01,r} \geq \gamma$, conventionally
              $\gamma = 3$ (substantial evidence) or $\gamma = 10$ (strong
              evidence).
        \item Perform a sensitivity analysis with respect to the prior
              distribution. For example, visualize the Bayes factors for different
              prior standard deviations to assess the robustness of the
              conclusions.
      \end{enumerate}
    \end{boxedminipage}
  \end{table}

\section{Conclusions}

The concept of ``replication success'' is inherently multifaceted. Reducing it
to a single criterion seems to be an oversimplification. Nevertheless, we
believe that the ``non-significance'' criterion -- declaring a replication as
successful if both the original and the replication study produce
non-significant results -- is not fit for purpose. This criterion does not
ensure that both studies provide evidence for the absence of an effect, it can
be easily achieved for any outcome if the studies have sufficiently small sample
sizes, and it does not control the relevant error rates. While it is important
to replicate original studies with null results, we believe that they should be
analyzed using more informative approaches.
Box~\hyperref[box:recommendations]{1} summarizes our recommendations.

Our reanalysis of the RPCB studies with original null results showed that for
most studies that meet the non-significance criterion, the conclusions are much
more ambiguous -- both with frequentist and Bayesian analyses. While the exact
success rate depends on the equivalence margin and the prior distribution, our
sensitivity analyses show that even with unrealistically liberal choices, the
success rate remains below 40\% which is substantially lower than the 73\%
success rate based on the non-significance criterion.

This is not unexpected, as a study typically requires larger sample sizes to
detect the absence of an effect than to detect its presence \citep[Section
11.5.3]{Matthews2006}. Of note, the RPCB sample sizes were chosen so that each
replication had at least 80\% power to detect the original effect estimate based
on a standard superiority test. However, the design of replication studies
should ideally align with the planned analysis \citep{Anderson2022} so if the
goal of the study is to ﬁnd evidence for the absence of an effect, the
replication sample size should be determined based on a test for equivalence,
see \citet{Flight2015} and \citet{Pawel2022c} for frequentist and Bayesian
approaches, respectively.

Our reanalysis of the RPP and EPRP studies with original null results showed
that Bayes factors indeed indicate some evidence for no effect in cases where
the non-significance criterion was satisfied, possibly due to the smaller
effects and typically larger sample sizes in these fields compared to cancer
biology. On the other hand, in most cases the precision of the effect estimates
was still limited so that only one study pair achieved replication success with
the equivalence testing approach. However, it is important to note that the
conclusions from the RPP and EPRP analyses are merely anecdotal, as there were
only four study pairs with original null results to analyze.

For both the equivalence test and the Bayes factor approach, it is critical that
the equivalence margin and the prior distribution are specified independently of
the data, ideally before the original and replication studies are conducted.
Typically, however, the original studies were designed to find evidence for the
presence of an effect, and the goal of replicating the ``null result'' was
formulated only after failure to do so. It is therefore important that margins
and prior distributions are motivated from historical data and/or field
conventions \citep{Campbell2021}, and that sensitivity analyses regarding their
choice are reported.

In addition, when analyzing a single pair of original and replication studies,
we recommend interpreting Bayes factors and TOST \textit{p}-values as
quantitative measures of evidence and discourage dichotomizing them into
``success'' or ``failure''. For example, two TOST \textit{p}-values
$p_{\mathrm{TOST}} =$ 0.049 and $p_{\mathrm{TOST}} =$ 0.051 carry similar
evidential weight regardless of one being slightly smaller and the other being
slightly larger than 0.05. On the other hand, when more than one pair of
original and replication studies are analyzed, dichotomization may be required
for computing an overall success rate. In this case, the rate may be computed
for different thresholds that correspond to qualitatively different levels of
evidence (e.g., 1, 3, and 10 for Bayes factors, or 0.05, 0.01, and 0.005 for
\textit{p}-values).

Researchers may also ask whether the equivalence test or the Bayes factor is
``better''. We believe that this is the wrong question to ask, because both
methods address different questions and are better in different senses; the
equivalence test is calibrated to have certain frequentist error rates, which
the Bayes factor is not. The Bayes factor, on the other hand, seems to be a more
natural measure of evidence as it treats the null and alternative hypotheses
symmetrically and represents the factor by which rational agents should update
their beliefs in light of the data. Replication success is ideally evaluated
along multiple dimensions, as nicely exemplified by the RPCB, EPRP, and RPP.
Replications that are successful on multiple criteria provide more convincing
support for the original finding, while replications that are successful on
fewer criteria require closer examination. Fortunately, the use of multiple
methods is already standard practice in replication assessment, so our proposal
to use both of them does not require a major paradigm shift.

While the equivalence test and the Bayes factor are two principled methods for
analyzing original and replication studies with null results, they are not the
only possible methods for doing so. A straightforward extension would be to
first synthesize the original and replication effect estimates with a
meta-analysis, and then apply the equivalence and Bayes factor tests to the
meta-analytic estimate similar to the meta-analytic non-significance criterion
used by the RPCB. This could potentially improve the power of the tests, but
consideration must be given to the threshold used for the
\textit{p}-values/Bayes factors, as naive use of the same thresholds as in the
standard approaches may make the tests too liberal \citep{Shun2005}.
Furthermore, there are various advanced methods for quantifying evidence for
absent effects which could potentially improve on the more basic approaches
considered here \citep{Lindley1998,Johnson2010,Morey2011,Kruschke2018,
  Stahel2021, Micheloud2022, Izbicki2023}.

\section{Acknowledgments}
We thank the RPCB, EPRP, and RPP contributors for their tremendous efforts and
for making their data publicly available. We thank Maya Mathur for helpful
advice on data preparation. We thank Benjamin Ineichen for helpful comments on
drafts of the manuscript. We thank the three reviewers and the reviewing editor
for useful comments that substantially improved the paper. Our acknowledgment of
these individuals does not imply their endorsement of our work. We thank the
Swiss National Science Foundation for financial support (grant
\href{https://data.snf.ch/grants/grant/189295}{\#189295}).

\section{Conflict of interest}
We declare no conflict of interest.

\section{Software and data}
The code and data to reproduce our analyses is openly available at
\url{https://gitlab.uzh.ch/samuel.pawel/rsAbsence}. A snapshot of the repository
at the time of writing is available at
\url{https://doi.org/10.5281/zenodo.7906792}. We used the statistical
programming language R version 4.3.2 \citep{R} for analyses. The R packages \texttt{ggplot2}
\citep{Wickham2016}, \texttt{dplyr} \citep{Wickham2022}, \texttt{knitr}
\citep{Xie2022}, and \texttt{reporttools} \citep{Rufibach2009} were used for
plotting, data preparation, dynamic reporting, and formatting, respectively. The
data from the RPCB were obtained by downloading the files from
\url{https://github.com/mayamathur/rpcb} (commit a1e0c63) and extracting the
relevant variables as indicated in the R script \texttt{preprocess-rpcb-data.R}
which is available in our git repository. The RPP and EPRP data were obtained
from the \texttt{RProjects} data set available in the R package
\texttt{ReplicationSuccess} \citep{Held2020}, see the package documentation
(\url{https://CRAN.R-project.org/package=ReplicationSuccess}) for details on
data extraction.

\section{Appendix A: Sensitivity analyses}

The post-hoc specification of equivalence margins $\Delta$ and prior
distribution for the SMD under the alternative $H_{1}$ is debatable. Commonly
used margins in clinical research are much more stringent \citep{Lange2005}; for
instance, in oncology, a margin of $\Delta = \log(1.3)$ is commonly used for log
odds/hazard ratios, whereas in bioequivalence studies a margin of
\mbox{$\Delta = \log(1.25) % = 0.22
  $} is the convention \citep[Chapter 22]{Senn2021}. These margins would
translate into margins of $\Delta = % \log(1.3)\sqrt{3}/\pi =
0.14$ and $\Delta = % \log(1.25)\sqrt{3}/\pi =
0.12$ on the SMD scale, respectively, using
the $\text{SMD} = (\surd{3} / \pi) \log\text{OR}$ conversion \citep[p.
233]{Cooper2019}. Similarly, for the Bayes factor we specified a normal
unit-information prior under the alternative while other normal priors with
smaller/larger standard deviations could have been considered. Here, we
therefore investigate the sensitivity of our conclusions with respect to these
parameters.

\begin{figure}[!htb]
  % \begin{fullwidth}
\begin{knitrout}
\definecolor{shadecolor}{rgb}{0.969, 0.969, 0.969}\color{fgcolor}
\includegraphics[width=\maxwidth]{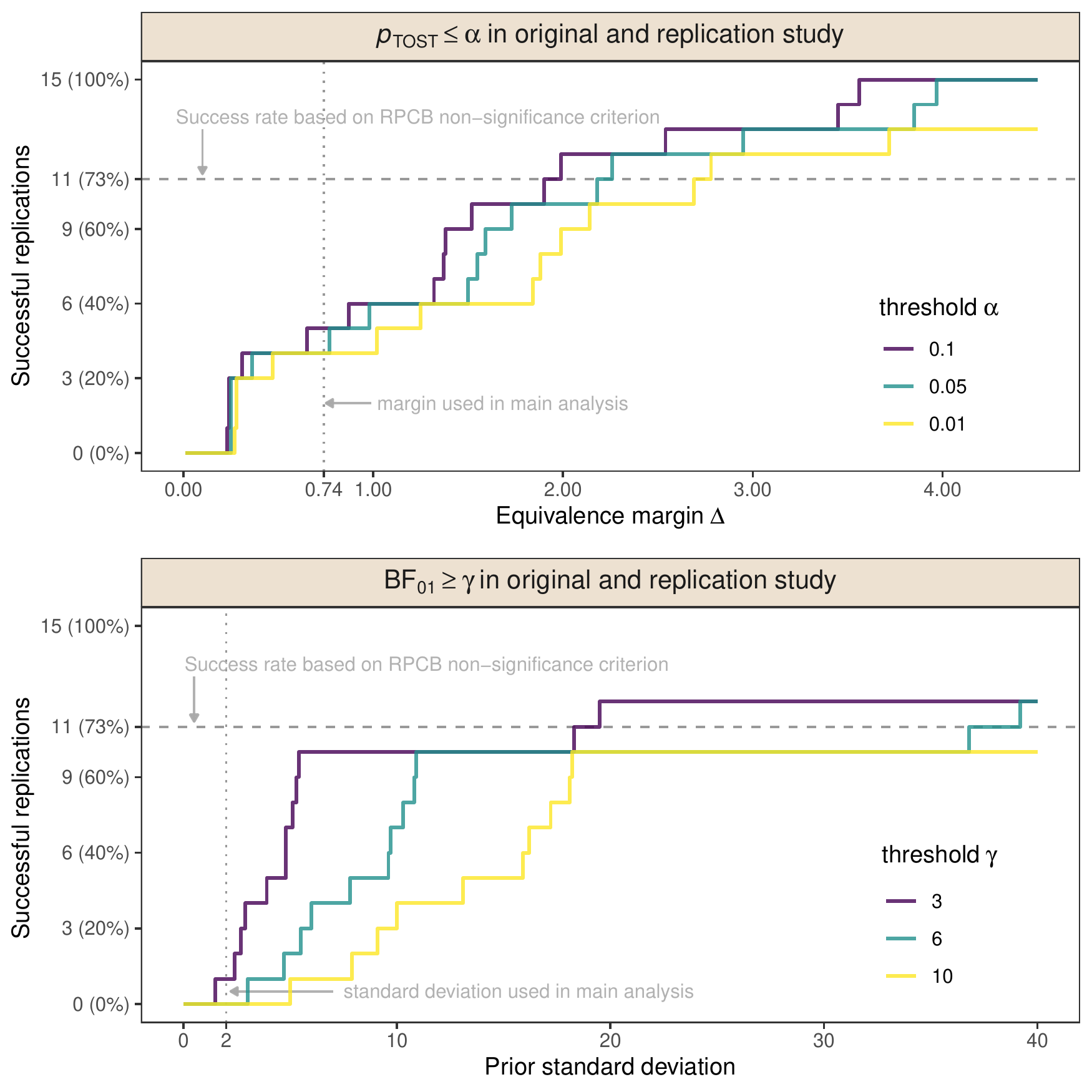} 
\end{knitrout}
\caption{Number of successful replications of original null results in the RPCB
  as a function of the margin $\Delta$ of the equivalence test
  ($p_{\text{TOST}} \leq \alpha$ in both studies for
  $\alpha = 0.1, 0.05, 0.01$) or the standard deviation of the zero-mean
  normal prior distribution for the SMD effect size under the alternative
  $H_{1}$ of the Bayes factor test ($\BF_{01} \geq \gamma$ in both studies for
  $\gamma = 3, 6, 10$).}
\label{fig:sensitivity}
% \end{fullwidth}
\end{figure}

The top plot of Figure~\ref{fig:sensitivity} shows the number of successful
replications as a function of the margin $\Delta$ and for different TOST
\textit{p}-value thresholds. Such an ``equivalence curve'' approach was first
proposed by \citet{Hauck1986}. We see that for realistic margins between $0$ and
$1$, the proportion of replication successes remains below $50\%$ for the
conventional $\alpha = 0.05$ level. To achieve a success rate of 11/15 =
73\%, as was achieved with the non-significance
criterion from the RPCB, unrealistic margins of $\Delta > 2$ are required.
Changing the success criterion to a more lenient level ($\alpha = 0.1$) or a
more stringent level ($\alpha = 0.01$) hardly changes the conclusion.

The bottom plot of Figure~\ref{fig:sensitivity} shows a sensitivity analysis
regarding the choice of the prior standard deviation and the Bayes factor
threshold. It is uncommon to specify prior standard deviations larger than the
unit-information standard deviation of $2$, as this corresponds to the
assumption of very large effect sizes under the alternatives. However, to
achieve replication success for a larger proportion of replications than the
observed 1/15 =
7\%, unreasonably large prior standard
deviations have to be specified. For instance, a standard deviation of roughly
$5$ is required to achieve replication success in $50\%$ of the replications at
a lenient Bayes factor threshold of $\gamma = 3$. The standard deviation needs
to be almost $20$ so that the same success rate 11/15 = 73\% as with the non-significance criterion is achieved. The necessary
standard deviations are even higher for stricter Bayes factor thresholds, such
as $\gamma = 6$ or $\gamma = 10$.

\section{Appendix B: Null results from the RPP and EPRP}
Here we perform equivalence test and Bayes factor analyses for the three
original null results from the Reproducibility Project: Psychology
\citep{Eastwick2008, Ranganath2008, Reynolds2008} and the original null result
from the Reproducibility Project: Experimental Philosophy \citep{McCann2005}. To
enable comparison of effect sizes across different studies, both the RPP and the
EPRP provided effect estimates as Fisher \textit{z}-transformed correlations
which we use in the following.

\begin{figure}[!htb]
\begin{knitrout}
\definecolor{shadecolor}{rgb}{0.969, 0.969, 0.969}\color{fgcolor}
\includegraphics[width=\maxwidth]{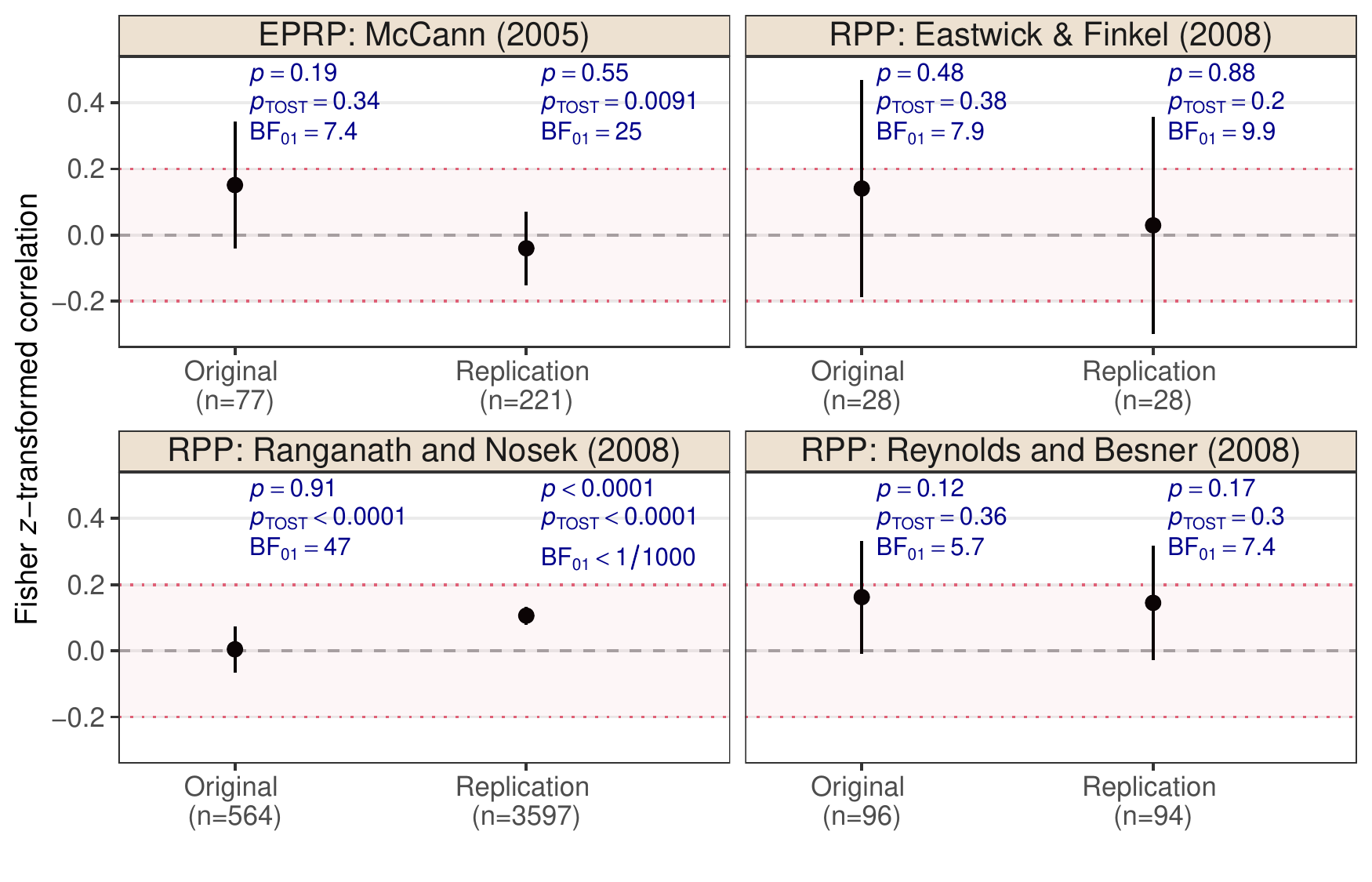} 
\end{knitrout}

\caption{Effect estimates on Fisher \textit{z}-transformed correlation scale
  with 90\% conﬁdence interval for the ``null results'' and their replication
  studies from the Reproducibility Project: Psychology
  \citep[RPP,][]{Opensc2015} and the Experimental Philosophy Replicability
  Project \citep[EPRP,][]{Cova2018}. The dashed gray line represents the value
  of no effect ($z = 0$), while the dotted red lines represent the equivalence
  range with a margin of $\Delta = 0.74$. The \textit{p}-value
  $p_{\text{TOST}}$ is the maximum of the two one-sided \textit{p}-values for
  the null hypotheses of the effect being greater/less than $+\Delta$ and
  $-\Delta$, respectively. The Bayes factor $\BF_{01}$ quantifies the evidence
  for the null hypothesis $H_{0} \colon z = 0$ against the alternative
  $H_{1} \colon z \neq 0$ with normal prior centered around zero and standard
  deviation of 2 assigned to the effect size under $H_{1}$.}
\label{fig:RPPandEPRPnull}
\end{figure}

Figure~\ref{fig:RPPandEPRPnull} shows effect estimates with
90\% confidence intervals, two-sided \textit{p}-values for
the null hypothesis that the effect size is zero, TOST \textit{p}-values for a
margin of $\Delta =$ 0.2, and Bayes factors using a normal prior
centered around zero with a standard deviation of 2. We see that
all replications except the replication of \citet{Ranganath2008} would be
considered successful with the non-significance criterion, as the original and
replication \textit{p}-values are greater than 0.05. In all three cases, the
Bayes factors also indicate substantial ($\BF_{01} >$ 3) to strong evidence
($\BF_{01} >$ 10) for the null hypothesis of no effect. Compared to the Bayes
factors in the RPCB, the evidence is stronger, possibly due to the mostly larger
sample sizes in the RPP and EPRP.

Interestingly, the opposite conclusion is reached when we analyze the data using
an equivalence test with a margin of $\Delta =$ 0.2 (which may be
considered liberal as it represents a small to medium effect based on the
\citealp{Cohen1992} convention). In this case, equivalence at the 5\% level can
only be established for the \citet{Ranganath2008} original study and its
replication simultaneously, as the confidence intervals from the other studies
are too wide to be included in the equivalence range. Furthermore, the
\citet{Ranganath2008} replication also illustrates the conceptual difference
between testing for an \emph{exactly zero} effect versus testing for an effect
\emph{within an interval around zero}. That is, the Bayes factor indicates no
evidence for a zero effect (because the estimate is clearly not zero), but the
equivalence test indicates evidence for a negligible effect (because the
estimate is clearly within the equivalence range).

\begin{figure}[!htb]
  % \begin{fullwidth}
\begin{knitrout}
\definecolor{shadecolor}{rgb}{0.969, 0.969, 0.969}\color{fgcolor}
\includegraphics[width=\maxwidth]{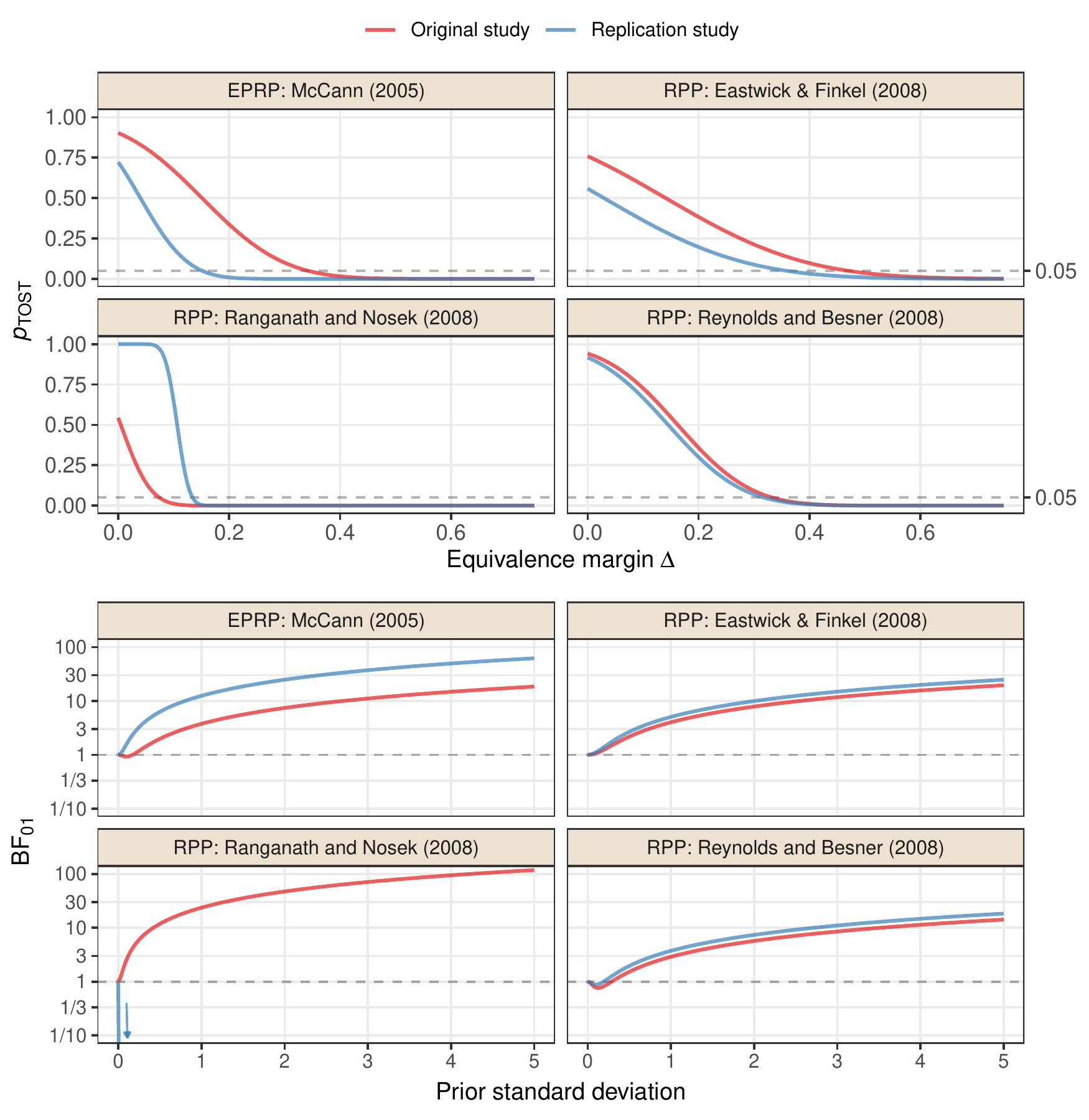} 
\end{knitrout}

\caption{Sensitivity analyses for the ``null results'' and their replication
  studies from the Reproducibility Project: Psychology
  \citep[RPP,][]{Opensc2015} and the Experimental Philosophy
     Replicability Project \citep[EPRP,][]{Cova2018}. The Bayes factor of the replication of
  \citet{Ranganath2008} decreases very quickly and is only shown for a limited
  range.}
\label{fig:sensitivityRPPandEPRP}
% \end{fullwidth}
\end{figure}

As before, the particular choices of the equivalence margin $\Delta$ for the
equivalence test and prior standard deviation of the Bayes factor are
debatable. We therefore report sensitivity analyses in
Figure~\ref{fig:sensitivityRPPandEPRP} which show the TOST \textit{p}-values and
Bayes factors of original and replication studies for a range of margins and
prior standard deviations, respectively. Apart from the \citet{Ranganath2008}
study pair, all studies require large margins of about $\Delta = 0.4$ to
establish replication success at the 5\% level (in the sense of original and
replication TOST \textit{p}-values being smaller than 0.05). On the other hand,
in all but the \citet{Ranganath2008} replication, the data provide substantial
evidence for a null effect ($\BF_{01} > 3$) for prior standard deviations of
about one, while larger prior standard deviations of about three are required
for the data to indicate strong evidence ($\BF_{01} > 10$) for a null effect,
whereas the data from the \citet{Ranganath2008} replication provide very strong
evidence against a null effect for all prior standard deviations considered.

\section{Appendix C: Technical details on Bayes factors}
We assume that effect estimates are normally distributed around an unknown
effect size $\theta$ with known variance equal to their squared standard error,
i.e.,
\begin{align*}
  \hat{\theta}_{i} \given \theta \sim \Nor(\theta, \sigma^{2}_{i})
\end{align*}
for original ($i = o$) and replication ($i = r$). This framework is similar to
meta-analysis and can be applied to many types of effect sizes and data
\citep[Section 2.4]{Spiegelhalter2004}. We want to quantify the evidence for the
null hypothesis that the effect size is equal to a null effect
($H_{0} \colon \theta = \theta_{0}$, typically $\theta_{0} = 0$) against the
alternative hypothesis that the effect size is non-null
($H_{1} \colon \theta \neq \theta_{0}$). This requires specification of a prior
distribution for the effect size under the alternative, and we will assume a
normal prior $\theta \given H_{1} \sim \Nor(m, s^{2})$ in the following. The
Bayes factor based on an effect estimate is then given by the ratio of its
likelihood under the null hypothesis to its marginal likelihood under the
alternative hypothesis, i.e.,
\begin{align*}
  \BF_{01,i}
  % &= \frac{\Pr(H_{0} \given \hat{\theta}_{i})}{\Pr(H_{1} \given \hat{\theta}_{i})} \, \bigg/ \,
  %   \frac{\Pr(H_{0})}{\Pr(H_{1})} \\
  &= \frac{p(\hat{\theta}_{i} \given H_{0})}{p(\hat{\theta}_{i} \given H_{1})} \\
  &= \frac{p(\hat{\theta}_{i} \given \theta_{0})}{\int_{-\infty}^{+\infty} p(\hat{\theta}_{i} \given \theta) \, p(\theta \given H_{1}) \, \mathrm{d}\theta} \\
  &= \sqrt{1 + \frac{s^{2}}{\sigma^{2}_{i}}} \, \exp\left[-\frac{1}{2} \left\{\frac{(\hat{\theta}_{i} -
              \theta_{0})^{2}}{\sigma^{2}_{i}} - \frac{(\hat{\theta}_{i} - m)^{2}}{\sigma^{2}_{i} + s^2}
              \right\}\right].
  % &= \frac{\Nor(\hat{\theta}_{i} \given \theta_{0}, \sigma^{2}_{i})}{\Nor(\hat{\theta}_{i}\given m, \sigma^{2}_{i} + v)}
\end{align*}
In the main analysis we used a normal unit-information prior, that is, a normal
distribution centered around the value of no effect ($m = 0$) with a standard
deviation $s$ corresponding to the standard error of an SMD estimate based on
one observation \citep{Kass1995b}. Assuming that the group means are normally
distributed \mbox{$\overline{X}_{1} \sim \Nor(\theta_{1}, 2\tau^{2}/n)$} and
\mbox{$\overline{X}_{2} \sim \Nor(\theta_{2}, 2\tau^{2}/n)$} with $n$ the total
sample size and $\tau$ the known data standard deviation, the distribution of
the SMD is
\mbox{$\text{SMD} = (\overline{X}_{1} - \overline{X}_{2})/\tau \sim \Nor\{(\theta_{1} - \theta_{2})/\tau, \sigma^{2} = 4/n\}$}.
The standard error $\sigma$ of the SMD based on one unit ($n = 1$), is hence
$2$, meaning that the standard deviation of the unit-information prior is
$s = 2$.
% , just as the unit standard deviation for log hazard/odds/rate ratio effect
% sizes \citep[Section 2.4]{Spiegelhalter2004}

\bibliography{bibliography}

\end{document}